\pdfoutput=1
\documentclass[twocolumn,superscriptaddress,aps,preprintnumbers,amsmath,amssymb,prl,nofootinbib]{revtex4-2}

\usepackage{amsmath}
\usepackage{amssymb}
\usepackage{amsfonts}
\usepackage{graphicx}
\usepackage{xcolor}
\usepackage{xfrac}
\usepackage{comment}
\usepackage{pifont}
\usepackage{physics}
\usepackage{fourier}
\usepackage{hyperref}
\usepackage{bm}
\usepackage{enumitem}

\definecolor{rossoferrari}{HTML}{D9073D}
\definecolor{mediumblue}{HTML}{0000CD}
\definecolor{forestgreen}{HTML}{228B22}
\definecolor{desy_blue}{HTML}{009EE2}
\definecolor{desy_orange}{HTML}{FD8800}
\definecolor{light_pink}{rgb}{1,0.4,0.4}
\definecolor{light_blue}{rgb}{0.284602,0.317763,0.963947}
\hypersetup{setpagesize=false,bookmarksnumbered=true,bookmarksopen=true,%
colorlinks=true,linkcolor=light_blue,urlcolor=rossoferrari,citecolor=rossoferrari,linktocpage=false}

\newcommand{\eV}{\,\mathrm{eV}}

\newcommand{\MeV}{\,\mathrm{MeV}}
\newcommand{\GeV}{\,\mathrm{GeV}}
\newcommand{\TeV}{\,\mathrm{TeV}}

\def\Mpl{M_\text{Pl}}

\newcommand{\lmk}{\left(}  
\newcommand{\rmk}{\right)}

\newcommand{\con}{{\rm (0)}}
\newcommand{\fo}{{\rm (1)}}
\newcommand{\so}{{\rm (2)}}

\def\beq#1\eeq{\begin{align}#1\end{align}}

\newcommand{\eq}[1]{Eq.~(\ref{#1})}

\begin{document}


\title{
Axion dark matter from first-order phase transition,\\ 
and very high energy photons from GRB 221009A
}

\author{Shota Nakagawa}
\email{shota.nakagawa.r7@dc.tohoku.ac.jp}
\affiliation{Department of Physics, Tohoku University, Sendai, Miyagi 980-8578, Japan}

\author{Fuminobu Takahashi}
\email{fumi@tohoku.ac.jp}
\affiliation{Department of Physics, Tohoku University, Sendai, Miyagi 980-8578, Japan}

\author{Masaki Yamada}
\email{m.yamada@tohoku.ac.jp}
\affiliation{Department of Physics, Tohoku University, Sendai, Miyagi 980-8578, Japan}
\affiliation{FRIS, Tohoku University, Sendai, Miyagi 980-8578, Japan}

\author{Wen Yin}
\email{yin.wen.b3@tohoku.ac.jp}
\affiliation{Department of Physics, Tohoku University, Sendai, Miyagi 980-8578, Japan}

\preprint{TU-1172}

\date{\today}


\begin{abstract}
\noindent
We study an axion-like particle (ALP) that experiences the first-order phase transition  with respect to its mass or potential minimum. 
This can be realized if the ALP obtains a potential from non-perturbative effects of SU($N$) gauge theory that is confined via the first-order phase transition, or if the ALP is trapped in a false vacuum at high temperatures until it starts to oscillate about the true minimum.
The resulting ALP abundance is significantly enhanced compared to the standard misalignment mechanism, explaining dark matter in a broader parameter space that is accessible to experiments e.g. IAXO, ALPS-II, and DM-radio. Furthermore, the viable parameter space includes a region of the mass $m_a \simeq 10^{-7} - 10^{-8}$ eV and the ALP-photon coupling $g_{a \gamma \gamma} \simeq 10^{-11} {\rm GeV}^{-1}$ that can explain the recent observation of very high energy photons from GRB221009A via axion-photon oscillations. The parameter region suggests that the FOPT can generate the gravitational wave that explains the NANOGrav hint.
If the ALP in this region explains dark matter, then the ALP has likely experienced a first-order phase transition.
\end{abstract}


\maketitle


\textit{\textbf{Introduction.\,---\,}} 
The Universe has experienced phase transitions in its thermal history as the temperature decreases by many orders of magnitude since the big bang. 
The physics of the phase transition can be universally understood by the behavior of the order parameter. 
For example, the second order phase transition (SOPT) is characterized by a critical exponent
which specifies the temperature dependence of the order parameter. 
The QCD phase transition occurring at a temperature around $100 \MeV$ is of this type. 
The electroweak phase transition proceeds via the crossover in the Standard Model (SM), where the order parameter changes smoothly. 
On the contrary, the order parameter changes discontinuously in the first-order phase transition (FOPT), which proceeds via the nucleation of true-vacuum bubbles. 
This is realized in many models for physics beyond the SM. 
The dynamics of the thermal Universe drastically changes if one considers a different order of phase transition. 
For example, 
the FOPT results in the production of gravitational waves (GWs) from the bubble collisions and the subsequent stochastic dynamics of the plasma. 
The production of topological defects, associated with the spontaneous symmetry breaking, is also modified qualitatively.

The QCD axion~\cite{Peccei:1977hh,Peccei:1977ur,Weinberg:1977ma,Wilczek:1977pj} and axion-like particles (ALPs) have been extensively studied in the literature 
as candidates for dark matter (DM)~\cite{Preskill:1982cy,Abbott:1982af,Dine:1982ah}, 
and their dynamics is strongly affected by the order of phase transitions. (See for review \cite{Jaeckel:2010ni,Ringwald:2012hr,Arias:2012az,Graham:2015ouw,Marsh:2015xka,Irastorza:2018dyq,DiLuzio:2020wdo}) 
The QCD axion acquires a potential from the non-perturbative effects of QCD. 
It is temperature-dependent during the QCD phase transition 
because the QCD phase transition is the SOPT. 
While these axions are expected to have constant mass from, for example, gravitational instanton effects, they may also acquire a temperature-dependent effective potential arising from instanton effects of thermalized hidden SU($N$) gauge sectors.
Then, depending on the order parameter of the confinement phase transition of SU($N$), the ALP may have a temperature-dependent  effective mass during the phase transition. 
In Ref.~\cite{Arias:2012az}, the authors considered the case in which the phase transition is the second order or a crossover, like the QCD. 
They derived the upper bound on the ALP abundance, which can be represented as a lower bound on the ALP decay constant to explain the DM density.

In this letter, we consider the case in which the ALP experiences the first-order phase transition. 
One of the examples is the first-order phase transition of SU($N$) confinement. 
The phase transition proceeds via the nucleation of true-vacuum bubbles. 
As the bubble goes through, the ALP potential suddenly grows within a very short time scale, and the ALP field value  does not change much during the phase transition. 
Another example is the so-called trapped misalignment mechanism~\cite{Higaki:2016yqk,DiLuzio:2021gos,Jeong:2022kdr}, where the axion is trapped in a false vacuum at high temperature and suddenly starts oscillating around the true vacuum when the potential barrier disappears. 
In the FOPT case, the resulting ALP abundance is significantly enhanced compared with the SOPT~\cite{Arias:2012az} or the standard misalignment mechanism~\cite{Preskill:1982cy,Abbott:1982af,Dine:1982ah}. This is because, in the case of FOPT, it is possible for the ALP to start oscillating with a large amplitude after its mass becomes much larger than the Hubble parameter.
Thus, the ALP produced in the FOPT
can explain DM for a broader parameter space which is more accessible to experiments such as  IAXO, ALPS-II, and DM-radio. 
We also discuss cosmological aspects of the dark sector that triggers the FOPT. In particular, we propose a possible solution to the cooling problem of dark glueballs.

Interestingly, 
the viable parameter space includes a region in which
the ALP-photon conversion can explain the observations of
very high energy photons from the extremely bright gamma-ray burst GRB 221009A by the Large High Altitude Air Shower Observatory (LHAASO) and Carpet-2.
 The GRB 221009A was detected by Fermi GBM and Swift~\cite{FermiGBM-GCN, Swift-GCN}, and
it was accompanied by $O(1000)$ gamma-rays observed by LHAASO with energy up to $18 \TeV$~\cite{LHAASO-GCN} and a photon-like airshower of 251 TeV by Carpet-2 ~\cite{CarpetATel-GRB}. Such very high energy photons could not have reached us from the reported redshift $z \simeq 0.151$~\cite{z-VLT-GCN,z-GTC-GCN,z-host-VLT-GCN} because of the efficient electron-positron pair creation within the extragalactic background radiation. 
If the observed very high energy gamma rays are not due to the Galactic source, this is a hint of the beyond SM physics. 
It was discussed that this apparent contradiction can be resolved if there exists an ALP with mass of order $10^{-7\,\text{-}\,8} \eV$ and the ALP-photon-photon coupling $g_{a \gamma \gamma}$ of order $10^{-11} \GeV^{-1}$~\cite{Galanti:2022pbg,Baktash:2022gnf,Troitsky:2022xso}. In this parameter region, the abundance of ALP produced by the ordinary misalignment mechanism is far below the observed DM abundance.\footnote{The ALP abundance is enhanced by the anharmonic effect if  the ALP initially sits near the hilltop by some dynamics~\cite{Daido:2017wwb,Takahashi:2019pqf, Co:2018mho, Takahashi:2019qmh, Kobayashi:2019eyg}. However, this direction does not work since the axionic isocurvature perturbation is significantly enhanced~\cite{Lyth:1991ub,Kobayashi:2013nva}. One exception is when the potential is very  flat near the hilltop. In this case the ALP slow-roll solution is an attractor, which suppresses the isocurvature perturbations~\cite{Nakagawa:2020eeg}. 
Another possibility is to make use of the clockwork mechanism to enhance the ALP coupling to photons~\cite{Farina:2016tgd,Higaki:2016yqk}, while keeping its oscillation amplitude large.}
We will see that, if the ALP is produced from the FOPT, it can naturally explain DM in this parameter region suggested by GRB 221009A. See Refs.~\cite{Li:2022wxc, Lin:2022ocj} for other phenomenological implications.

We also discuss the formation of quasi-stable non-topological solitons called oscillons after the phase transition. 
The formation of oscillon requires an $\mathcal{O}(1)$ density perturbations at the phase transition for the case of the SOPT, including the case for QCD axion. This is because the amplitude of ALP changes adiabatically and its perturbations cannot grow much during the SOPT. 
This is in contrast to the case of FOPT, where the amplitude of ALP oscillation is as large as its decay constant and the duration of the instability is long enough for its perturbations to grow exponentially after the phase transition. 
In other words, the scenario with the FOPT inevitably results in the formation of oscillons.


\smallskip
\textit{\textbf{ALPs from sudden change of potential.\,---\,}} 
In this letter, we denote the ALP mass as $m_a(T)$ and its present value as $m_0$. 
In the standard scenario, it is usually assumed that the ALP has a constant mass ($m_a(T) = m_0$) from, e.g., a gravitational instanton effect or other explicit symmetry breaking.
The ALP abundance in this case can be calculated as\footnote{Here and in what follows,
the upper indices $(0),(1),(2)$ mean that the variables are for cases with the standard constant mass, FOPT and SOPT, respectively. 
} 
\beq
 \frac{\rho_a^\con}{s} 
 \simeq 
 \frac{45 c^{3/2} m_0^{1/2} a_i^2}{4 \pi^2 g_{*s}^\con \Mpl^{3/2}} 
 \lmk \frac{\pi^2 g_*^\con}{90} \rmk^{3/4}, 
 \label{eq:const}
\eeq
where $a_i$ is the initial ALP field value, $g_*$ ($g_{*s}$) is the effective relativistic degrees of freedom for energy (entropy) density, and $\Mpl \simeq 2.4 \times 10^{18}$ GeV is the reduced Planck mass. The most natural value of $a_i$ is of order the decay constant, $f_a$. Here 
$g_*^\con$  and $g_{*s}^\con$ are evaluated at $T=T_{\rm osc}^\con$, where
the oscillation temperature $T_{\rm osc}^\con$ is 
determined by $c H (T_{\rm osc}^\con) = m_0$ with $c \simeq 3$ being a numerical constant. 
This gives 
\beq
 T_{\rm osc}^\con \simeq \lmk \frac{m_0 \Mpl}{c}\sqrt{\frac{90}{\pi^2 g_*^\con}} \rmk^{1/2}.
\eeq

The ALP can have a Chern-Simons coupling to a hidden SU($N$) gauge field, in which case it acquires a potential via the instanton effect. 
If the confinement phase transition of SU($N$) is of the second order, 
the ALP has the temperature-dependent potential such as 
\beq
 m_a(T) = m_0 \lmk \frac{T}{T_c} \rmk^{-n/2}, 
\eeq
where $T_c$ is a temperature at which the ALP mass becomes as large as the present value and $n$ is a critical exponent of the phase transition. 
According to the dilute instanton gas approximation, the exponent is given by $n = 11 N/3 + N_f/3-4$ for SU($N$) with $N_F$ flavors, which is consistent with lattice calculations in several examples. 
We assume that the ALP mass becomes constant at $T  = T_c$ for simplicity, though this does not affect our conclusions. 
In this case, 
the ALP starts to oscillate at $c H (T_{\rm osc}^\so) = m(T_{\rm osc}^\so)$. 
Subsequently, the ALP mass changes adiabatically, 
$\abs{\dot{m}/ m^2} \simeq (n/4)H(T)/m(T) \ll 1$ for $T \ll T_{\rm osc}^\so$, 
and its number density per entropy density becomes almost constant. 
Its abundance at present is then given by 
\beq
 \frac{\rho_a^\so}{s} 
 &\simeq
 \lmk \frac{m_0}{m_a (T_{\rm osc}^\so)} \rmk^{1/2} \times \left. \frac{\rho_a^\con}{s} \right\vert_{ 
 g_{*(s)}^\con \to g_{*(s)}^\so
 }
 \\
 &\simeq 
 \frac{45 a_i^2}{4 \pi^2 g_{*s}^\so} \frac{m_0^{(n+2)/(n+4)}}{T_c^{n/(n+4)}} 
 \lmk \frac{c'}{\Mpl} \sqrt{\frac{\pi^2 g_*^\so}{90}} \rmk^{(n+6)/(n+4)},
 \label{eq:2nd}
\eeq
where $g_{*(s)}^\so = g_{*(s)}(T_{\rm osc}^\so)$.
For example, the QCD phase transition is the second order and its exponent is $n \simeq 7.84$ with $T_c \simeq 147 \MeV$ for the QCD axion, which is obtained by fitting the results of lattice simulations~\cite{Borsanyi:2016ksw} up to higher temperature~\cite{Nakagawa:2020zjr}.

Now let us consider the case in which the ALP mass (or its potential minimum) changes instantaneously via the FOPT. 
For definiteness, we consider a simplified case with the sudden change from negligibly small axion mass to a constant mass $m_0$.
In this case, the result is similar to the case with the constant mass but with independent parameters $m_0$ and $T_{\rm osc}^\fo$: 
\beq
\frac{\rho_a^\fo}{s} 
 \simeq 
 \frac{45 m_0^2 a_i^2}{4 \pi^2 g_{*s}^\fo \lmk T_{\rm osc}^\fo \rmk^{3}},
 \label{eq:1st}
\eeq
where $T_{\rm osc}^\fo$ is now determined by the temperature at the FOPT.  
Here we have assumed $c H (T_{\rm osc}^\fo) < m_0$. 
If this is not satisfied, 
the ALP does not start to oscillate at the FOPT. 
The scenario in this case is similar to the case with the constant mass and the abundance is given by \eq{eq:const}. Note that the numerator in Eq.~(\ref{eq:1st}) is determined by the potential height at the onset of oscillations, and it is independent of $f_a$ in the case of the QCD axion, which enables the QCD axion near the lower end of the axion window to explain DM~\cite{Higaki:2016yqk,Jeong:2022kdr}.

We are interested in the case in which the ALP is DM. 
Then it should start to oscillate well before the matter-radiation equality. 
According to Ref.~\cite{Sarkar:2014bca} (see also Ref.~\cite{Das:2020nwc}), the lower bound on the oscillation temperature is 
\beq
\label{eq:bound}
 T_{\rm osc}^{(i)}  \gtrsim (1+z_{\rm DM})  T_0, 
\eeq
with $z_{\rm DM} \sim 10^6$ for $i=0,1,2$, 
where $T_0$ ($\simeq 2.3 \times 10^{-4} \eV$) is the temperature at present. 
This gives an upper bound on the ALP abundance for a given $m_0$ and $a_i$.
In the case of SOPT, 
the ALP mass should also be constant by that time, $T_c \gtrsim (1+z_{\rm DM})  T_0$, because the equation of state for DM should not largely deviate from $0$ after the redshift of $z_{\rm DM}$~\cite{Sarkar:2014bca}.


\begin{figure}[t]
	\centering
 	\includegraphics[width=1.0\linewidth]{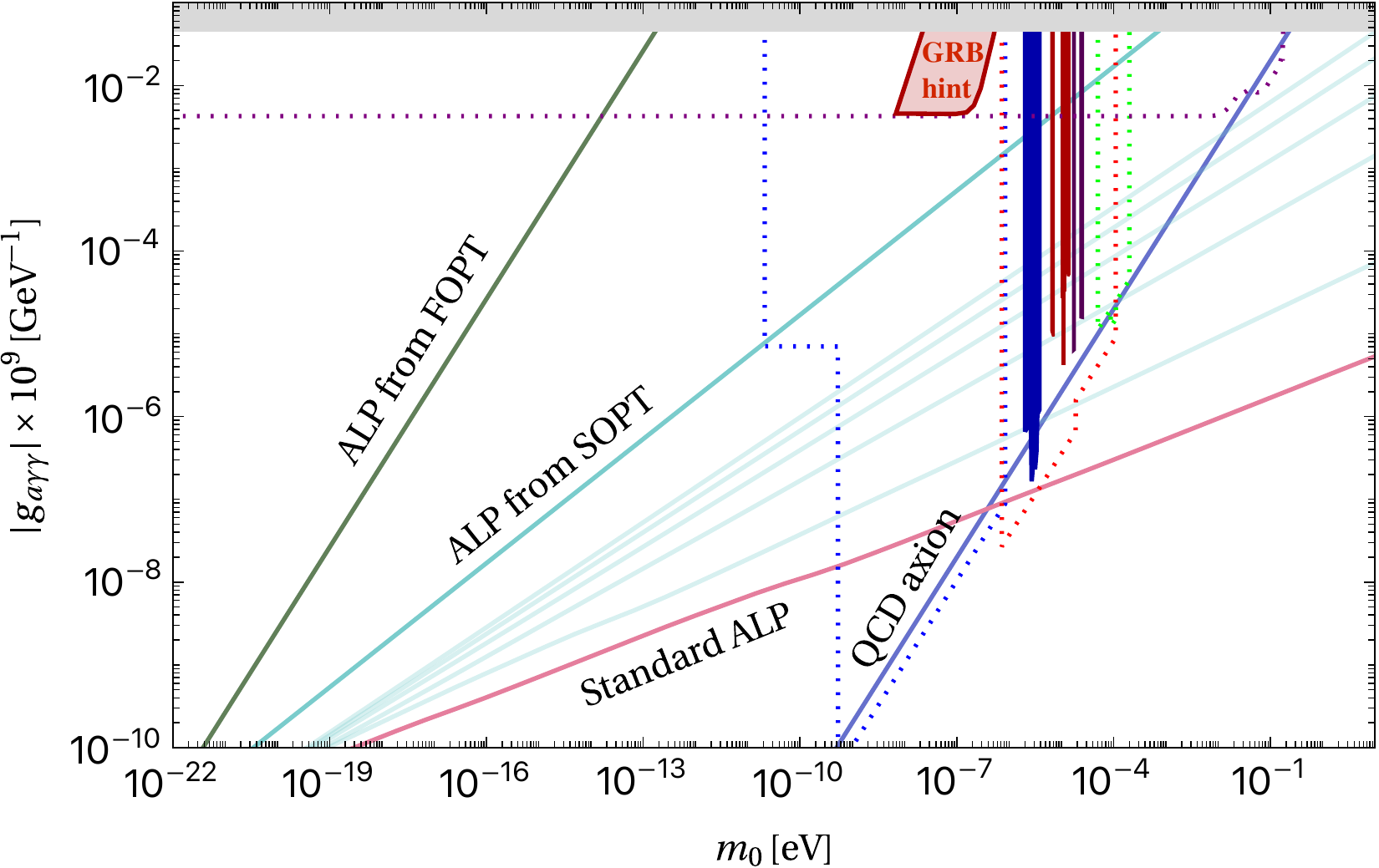} 
	\caption{ 
	Upper bounds on $\abs{g_{a\gamma\gamma}} = \alpha / (2 \pi f_a)$ 
	for the case with the constant mass (magenta), SOPT (cyan), and FOPT (green) to explain the DM abundance. The blue line represents the $\abs{g_{a\gamma\gamma}}-m_0$ relation for a typical QCD axion. 
	The thin lines (cyan) represent the upper bound for a fixed $n$ ($=1,3,5,7,9$ from bottom to top). 
	The GRB221008A hint can be explained in the red shaded region \cite{Baktash:2022gnf,Troitsky:2022xso}. 
Gray, blue, red, and purple shaded regions show the existing bounds from Globular clusters, ADMX, CAPP, and HAYSTAC. Purple, blue, red, and green dotted lines denote future sensitivities for IAXO, DM-radio, ADMX, and MADMAX, respectively. 
    }
	\label{fig:1}
\end{figure}


The observed DM density is given by $\rho_{\rm DM} / s \simeq 0.44 \eV$. 
Figure~\ref{fig:1} shows the upper bound on $\abs{g_{a\gamma\gamma}}$ below which the ALP can explain all DM for the cases with the constant mass (magenta line, \eq{eq:const}), SOPT (cyan, \eq{eq:2nd}), and FOPT (green line, \eq{eq:1st}), where $g_{a \gamma\gamma} = \alpha/(2 \pi f_a)$ {with $\alpha=1/137$ being the fine-structure constant}. 
We take $a_i = f_a$ 
and neglect the anharmonic effect for simplicity. 
The upper bound for the case with FOPT is calculated by using \eq{eq:bound}. 
The blue line represents the mass-decay-constant relation for the QCD axion, 
where 
$m_{{\rm QCD},a} \simeq 5.7 \mu \, {\rm eV} (f_a/10^{12} \GeV)^{-1}$ \cite{GrillidiCortona:2015jxo,Gorghetto:2018ocs}. 

We also plot the upper bounds for the case with SOPT with a fixed value of $n = 1,3,5,7,9$ as the thin lines (cyan) from bottom to top~\cite{Arias:2012az}, where we used $T_c \gtrsim (1+z_{\rm DM}) T_0$. The cyan line corresponds to the case with $n \to \infty$, which however is not consistent with the adiabatic condition. 
The region above the cyan line and below the green line is a new parameter space where the ALP can explain DM only in the case of FOPT.

The red shaded region is the favored region to explain the high energy photons from GRB 221009A from axion-photon conversion~\cite{Baktash:2022gnf,Troitsky:2022xso}.  
Gray, blue, red, and purple shaded regions show the constraints from Globular clusters~\cite{Dolan:2022kul} (see also Refs \cite{Raffelt:1985nk,Raffelt:1987yu,Raffelt:1996wa, Ayala:2014pea, Straniero:2015nvc, Giannotti:2015kwo, Carenza:2020zil}), ADMX~\cite{ADMX:1998pbl}, CAPP~\cite{Semertzidis:2019gkj, Lee:2020cfj}, and HAYSTAC~\cite{Brubaker:2016ktl}. Purple, blue, red, and green dotted lines denote the future sensitivities by IAXO~\cite{Irastorza:2011gs, Armengaud:2014gea, Armengaud:2019uso, Abeln:2020ywv} (see also ALPS II reach~\cite{Ortiz:2020tgs}),  DM-radio~\cite{DMRadio:2022pkf}, ADMX~\cite{Stern:2016bbw}, and MADMAX~\cite{Beurthey:2020yuq}, respectively. 
More complete bounds can be found in
Refs.\,\cite{AxionLimits, Semertzidis:2021rxs}, from which we adopted typical bounds and reaches. We also note some of the astrophysics bounds, such as the polarization measurement of magnetic white dwarfs~\cite{Dessert:2022yqq}, is in mild tension with the GRB hint.

In the following two sections, we provide a couple of models that realize the sudden change of the ALP potential.


\smallskip
\textit{\textbf{Model with a trapped potential.\,---\,}} 
We first consider a model in which the potential minimum of ALP changes via the FOPT. 
Suppose that the ALP potentail receives two different contributions, one of which has a temperature dependence: 
\beq
 V(a) = V_0(a) + V_T(a, T), 
\eeq
where $V_0$ and $V_T$ represent the sine-Gordon functions with a constant or temperature-pendent potential height, respectively. The periodicity of both terms are expected to be of the order of $2\pi f_a$ in a simple model. 

Two possibilities have been studied in the literature, among others.
The first case is that 
the ALP is initially trapped by $V_T(a, T)$ that becomes weaker at  lower temperatures and the potential becomes eventually dominated by the constant potential $V_0$. That is, $V_T \gg V_0$ at high $T$ whereas $V_T \ll V_0$ at low $T$. 
Such a time-dependent potential can be realized by the Witten effect \cite{Witten:1979ey,Fischler:1983sc} in the presence of (hidden) monopoles in the plasma~\cite{Kawasaki:2015lpf,Nomura:2015xil,Sato:2018nqy,Nakagawa:2020zjr}. 
The second case is that 
the ALP is trapped by the constant potential $V_0$ at high temperatures and then the potential becomes dominated by the temperature-dependent potential $V_T$ at low temperatures. 
That is, $V_T \ll V_0$ at high $T$ whereas $V_T \gg V_0$ at low $T$. 
This is the case, e.g., an ALP has a constant potential as well as a potential from non-perturbative effects of the gauge interactions. Such a possibility was considered in Refs.~\cite{Higaki:2016yqk,DiLuzio:2021gos,Jeong:2022kdr} in the context of QCD axion or some extension.

In both cases, 
one can consider the case in which the 
ALP is trapped at a false vacuum at high temperatures 
and then tunnels to the true vacuum at a certain temperature. 
Let us define the critical temperature as the temperature when the false vacuum becomes unstable. 
We also denote the inflection point of the potential at the critical temperature (or the field value at the false vacuum just before the critical temperature) as the critical value $a_*$.

The tunneling is described by an $O(4)$-symmetric instanton $a(r)$ obeying 
\beq
 \frac{d^2 a}{d r^2} + \frac{3}{r} \frac{da}{dr} = \frac{dV}{da}, 
\eeq
where $r$ represents the radius in $O(4)$ symmetric spacetime.
The tunneling rate is determined by the Euclidean instanton action $S_E$ such as $P \propto e^{-S_E}$, where 
\beq
 S_E = \int_0^\infty 2 \pi^2 r^3 \lmk \frac12 \lmk \frac{da}{dr} \rmk^2 + V(a(r)) - V_{\rm FV} \rmk,
\eeq
where $V_{\rm FV}$ is the potential height at the false vacuum. 
If we rescale $a$, $V$, and $r$ by $f_a$, $m_0^2 f_a^2$, and $m_0^{-1}$, respectively, 
the Euclidean instanton action scales such as $S_E \propto f_a^2 / m_0^2$ ($\gg 1$). 
This implies that the tunneling point must be extremely close to the false vacuum to have a reasonably small $S_E$ or a sufficiently large tunneling rate. 
This is possible just before the critical temperature when the false vacuum becomes unstable. 
Therefore we can approximate the amplitude of the axion oscillation after the tunneling as $a_*$. 
Then the ALP starts to oscillate around the true vacuum after the tunneling. 
The resulting abundance of ALP is given by \eq{eq:1st} with $T_{\rm osc}^{(1)}$ given by the temperature of the phase transition and $a_i$ by $a_*$. 
We expect that $a_* \sim f_a$ without fine-tuning. On the other hand, the isocurvature perturbation of the ALP tends to be suppressed in the trapped scenario~\cite{Kawasaki:2015lpf,Nomura:2015xil,Kawasaki:2017xwt,Jeong:2022kdr}.


\smallskip
\textit{\textbf{Model with a first-order phase transition.\,---\,}} 
Next, we provide a model in which the ALP obtains a potential
via the FOPT in a gauge sector. 
Suppose that an ALP has a Chern-Simons coupling to the dark SU($N$) gauge sector. We  consider the potential from the non-perturbative effect of the dark SU($N$) at IR. 
When the flavor number $N_F$ for the SU($N$) gauge theory is small enough, we expect that the deconfinement/confinement phase transition of SU($N$) gauge interaction proceeds via the FOPT~\cite{Lucini:2005vg}. 
The ALP obtains the effective potential via the instanton effect, which becomes strong at the SU($N$) confinement phase transition. 
This model thus realizes the scenario in which the ALP mass (or potential) experiences the FOPT (see e.g. \cite{Cui:2022vsr}). 
As the ALP mass changes instantaneously via the FOPT, 
its field value does not change during the phase transition. 
Then we can use \eq{eq:1st} for the ALP abundance given the relation between $T$ and  the temperature of the SU($N$) sector, $T_{\rm dark}$. This temperature depends on the energy density of dark gauge bosons produced in the UV theory, $\rho_{\rm dark}$, via e.g. the inflaton coupling and/or higher dimensional couplings between the SU($N$) and the SM sector. We have $T_{\rm dark}\sim \rho^{1/4}_{\rm dark}$ due to the fast thermalization of the SU($N$) gauge interaction.\footnote{
The axion radiation produced from the SM particles could produce the SU($N$) gauge bosons. This contribution is not sizable unless the temperature is close to the decay constant, because the production of the dark gauge bosons in the high energy is either through the ALP decay, whose rate is suppressed by the small axion mass and boost factor, or scattering between/mediated by ALP whose rate is suppressed by $1/f_a^4.$ 
}
Instead of specifying the UV setup for the production of the dark sector, we simply use the temperature ratio, $\eta =\left. T_{\rm dark}/T \right\vert_{T = T_{\rm osc}}$,  to parameterize its abundance. Then, we obtain the correct ALP abundance for DM, if
\beq 
g_{a\gamma\gamma}\sim 8\times 10^{-11} {\rm GeV}^{-1} c_\gamma \lmk \frac{m_0}{10^{-8}\,{\rm eV}} \rmk  \left(\frac{\eta}{0.05}\right)^6 \left(\frac{a_i}{f_a}\right)^4 \,.
\eeq 
Here we used that the dark sector confines at $T_{\rm dark}\sim \Lambda \sim \sqrt{m_0 f_a}$ and we used Eq.\,\eqref{eq:1st} for the axion abundance. Here we recovered the factor $c_\gamma$, which was taken to be $1$ in the previous analysis, as   $g_{a\gamma\gamma}\equiv c_\gamma \alpha /(2\pi f_a)$.

\smallskip
\textit{\textbf{Cooling of the dark sector.\,---\,}}
In the second model, the dark sector glueballs remain to contribute to the abundance of DM, whose temperature remains hot and marginally relativistic unless a further entropy is released. This is well known in the strongly interacting dark matter scenario ~\cite{Carlson:1992fn} (see also \cite{Dolgov:1980uu,Dolgov:2017ujf}). 
The energy density of glueballs can be approximated as the confinement scale $\Lambda^4$ at the confinement, and it approximately scales as matter (with a logarithmic correction) due to the cosmic expansion.
The contribution of the glueball density may be comparable to or slightly larger than the ALP abundance depending on $a_i$ and the $N$ if there were no extra entropy release. 

This problem, which also exists in the SOPT case, can be solved if $f_a$ is small enough because the dark sector entropy is released to the ALP radiation via, e.g., two CP-even glueballs $\to$ CP-odd glueball $+a$ (see the spectrum of the glueball Ref.\,\cite{Forestell:2016qhc}) followed by the CP-odd glueball $\to a$ + CP-even glueball in the case of pure SU($N$) theory. These reactions cool down the dark sector and suppress the glueball abundance. 
The interaction rate is $\Gamma_{\rm cool}\sim  C_{\rm cool}\frac{n_{\rm gb}}{f^2_a}$ at the confinement according to the na\"{i}ve dimensional analysis with $C_{\rm cool}$ being the numerical factor representing the theoretical uncertainty, and $n_{\rm gb}$ the number density of the glueballs. 
If  $ \Gamma_{\rm cool}\gg H \label{cool}$ at the phase transition, the dark sector efficiently loses its energy 
until $n_{\rm gb}$ decreases to satisfy
\beq
\Gamma_{\rm cool}\sim H
\eeq
like the freeze out of WIMP. 
By requiring the final glueball abundance to be below the ALP abundance, 
we obtain $g_{a\gamma\gamma}\gtrsim 2\times 10^{-10}\,{\rm GeV}^{-1} \frac{c_\gamma}{C_{\rm cool}^{6/11}m_0^{1/11}{(a_i/f_a)}^{4/11}}$. 
Interestingly, the lower bound is close to the regime that explains the very high energy photons from the GRB 221009A via the axion-photon oscillation and can be probed in future experiments for the solar axion and dark matter detection depending on the numerical factors. Even if $C_{\rm cool}\lesssim 1$ is not very large, we can have the allowed region by taking $c_\gamma \lesssim 0.01$. This mechanism predicts dark radiation of the ALP as a remnant of the cooling of dark sector. When the ALP initial misalignment angle is small, which is the case for the low scale inflation~\cite{Graham:2018jyp, Guth:2018hsa, Ho:2019ayl}, $\eta$ can be relatively large, inducing the sizable dark radiation that can be probed in the future. In this case, we have sizable dark sector energy for the phase transition leading to sufficient GWs from the FOPT. 

Alternatively, we can introduce other particles/interactions to solve this problem. For instance, we can introduce dark quarks with dark photon couplings. It does not suppress the ALP mass much if the dark quark mass is not extremely small compared to $\Lambda$, but the dark pion can decay or annihilate into the dark photons to cool the dark sector. 
Detecting those light particles can also probe our scenario. 
The prediction of dark radiation and GWs is universal for the FOPT if $T_{\rm dark}\sim T$ before the phase transition. In the light quark scenario we can have the natural $a_i\approx f_a$ while suppressing $m_0 <\Lambda^2/f_a$.
The ALP abundance suggests $T^{(1)}_{\rm osc}$ around MeV for $m_{0}= 10^{-(7\text{-}8)}\,{\rm eV}$ and  $ f_a=10^{8-9}$GeV. The parameter region, therefore, can simultaneously explain the NANOGrav hint for the GWs (see e.g. Ref.~\cite{Nakai:2020oit}).

A similar cooling problem may arise in the first model with the trapping potential. We expect that it can also be resolved by a similar mechanism.


\smallskip
\textit{\textbf{Growth of perturbations and oscillon formation.\,---\,}} 
Finally, we comment on the growth of perturbations for the ALPs and discuss the formation of oscillons using the linear perturbation theory. 

For simplicity, we approximate 
the ALP potential by the polynomial potential at the onset of oscillation such as 
\beq
 V(a) 
 &= \Lambda^4 \lmk 1-\cos \frac{a}{f_a} \rmk 
 \\
 &\simeq \Lambda^4 \lmk \frac{a^2}{2 f_a^2} 
 - \frac{a^4}{24 f_a^4} + \frac{a^6}{720 f_a^6} \rmk, 
\eeq
where $m_0 \equiv \Lambda^2 / f_a$. 
We decompose the ALP as a homogeneous part and a perturbation from it: 
$a = a(t) + \delta a(x,t)$. 
Assuming that $\delta a(x,t)$ is much smaller than the time-averaged value of $a(t)$, we can use the linear perturbation theory 
to calculate its growth rate $\mu (k)$, where $\delta a(k,t) \propto e^{\mu(k) t}$ after the Fourier transformation. 
The result is given by 
\beq
 \mu(k) \simeq \frac{k}{2} 
 \sqrt{\lmk \frac{a}{2f_a} \rmk^2 \lmk 1 - \frac{a^2}{12 f_a^2} \rmk 
 - \frac{k^2}{m_0^2} }, 
\eeq
for $k / m_0 \ll 1$ (see, e.g., Refs.~\cite{Kusenko:1997si,Doddato:2011fz}). 
This means that the growth rate is $\mathcal{O}(1)m_0$ for $a / f_a = \mathcal{O}(1)$. 
Then, if the amplitude of ALP decreases during the phase transition, $\mu(k)$ soon becomes to be complex and the perturbations cannot grow. This is the case for the SOPT. 
For the case with a constant ALP mass, 
the amplitude decreases due to the cosmic expansion for the time scale of $1/H$. Since $H_{\rm osc} \sim m_0$, there is no enough time for perturbations to grow by many orders of magnitude in this case. 
On the contrary, for the case of FOPT, the amplitude does not change much during the FOPT, and the time scale of growth, $1/m_0$, is much shorter than the Hubble expansion rate, $1/H$. Therefore the perturbations exponentially grow after the onset of the oscillation of ALP. 
This results in the formation of oscillons, which are the quasi-stable non-topological solitons in real scalar field theories. 
This is a characteristic consequence of our scenario compared with the other two scenarios.

Below the green line and above the cyan line in Fig.~\ref{fig:1}, 
the ALP must be produced via our mechanism. 
If one detects an ALP in this parameter space, one can conclude that 
the oscillons form at the phase transition. 
The oscillons are quasi-stable but decay in the cosmological time scale after the formation.


\smallskip
\textit{\textbf{Discussion.}}\,---\,In this letter, we have provided a new scenario for the production of coherent oscillation of ALP DM, motivated by the time-dependent ALP mass or potential minimum. 
Contrary to the standard scenario, in which the ALP mass and potential minimum are constant or experience a SOPT, we have considered the case with a FOPT (or a sudden change) in terms of the ALP mass or potential minimum. 
Possible models for realizing such FOPT have also been proposed. 
Our scenario provides a broader parameter space in which the ALP can explain all DM. Interestingly, the recently reported hint of ALP by gamma-ray burst from GRB 221009A can be DM only if it is produced by our mechanism.

Our scenario is testable because we predict that the ALP is DM, namely, the ALP exists in the present Universe. In particular, the DM-radio can search for the ALP DM in some parameter of interest, particularly in the parameter space that can explain the GRB hint. 
If it observes ALP DM in the parameter space between the cyan and green lines in Fig.~\ref{fig:1}, it must be produced by our mechanism.

GWs can be generated during the formation of oscillons because the density perturbations for the ALP become as large as $\mathcal{O}(1)$. Unfortunately, the amplitude of GWs is too small to be observed by GW experiments in the near future. 
GWs are produced also from the bubble collisions and the stochastic dynamics of ambient plasma after the FOPT. The peak frequency of this GW spectrum is expected to be $\mathcal{O}(1) H^{-1}_{\rm form}$ with the redshift factor. The resulting peak frequency at present is too small to be detected by any observations in the parameters of interest. Note, however, that the GW from the FOPT of SU($N$) can be sizable and explain the NANOGrav hint and be probed in the future, if the dark sector is cooled by the ALP production or other exended dark sectors.

If there are both the QCD axion and the ALP, the level crossing could take place during the QCD phase transition~\cite{Hill:1988bu,Kitajima:2014xla,Daido:2015bva,Daido:2015cba,Ho:2018qur}. Then, the ALP produced before the QCD phase transition will be converted to the QCD axion with a heavier mass. Therefore, if this is combined by the FOPT or SOPT scenarios, the total abundance of the QCD axion and ALP can be further enhanced.


\medskip\noindent\textit{Acknowledgments.}\,---\,
The present work is supported by the Graduate Program on Physics for the Universe of Tohoku University (S.N.), JST SPRING, Grant Number JPMJSP2114 (S.N.), JSPS KAKENHI Grant Numbers  20H01894 (F.T.) 20H05851 (F.T., W.Y., and M.Y.), 21K13910 (M.Y.), 21K20364 (W.Y.), 22K14029 (W.Y.), and 22H01215 (W.Y.). MY was supported by MEXT Leading Initiative for Excellent Young Researchers.


\bibliographystyle{JHEP}
\bibliography{ref}


\end{document}